\documentstyle[12pt]{article}
\input{epsf}
\catcode`\@=11

\@addtoreset{equation}{section}
\def\theequation{\arabic{section}.\arabic{equation}}

\catcode`\@=11
\def\section{\@startsection{section}{1}{\z@}{3.5ex plus 1ex minus
   .2ex}{2.3ex plus .2ex}{\large\bf}}

\def\eqnarray{\let\@currentlabel=
\theequation\refstepcounter{equation}
    \global\@eqnswtrue
    \global\@eqcnt\z@\tabskip\@centering\let\\=\@eqncr
    $$\halign to \displaywidth\bgroup\@eqnsel\hskip\@centering
      $\displaystyle\tabskip\z@{##}$&\global\@eqcnt\@ne
       \hfil${{}##{}}$\hfil
      &\global\@eqcnt\tw@ $\displaystyle\tabskip\z@{##}$\hfil
       \tabskip\@centering&\llap{##}\tabskip\z@\cr}
\def\lefteqn#1{\hbox to 4\arraycolsep{$\displaystyle #1$\hss}}
\def\thesection{\arabic{section}.}

\def\appendix{\setcounter{section}{0}
        \def\thesection{Appendix.}
        \def\theequation{\Alph{section}.\arabic{equation}}}

\long\def\@makefntext#1{\parindent 0cm\noindent
\hbox to 1em{\hss$^{\@thefnmark}$}#1}

\def\IR{{\hbox{{\rm I}\kern-.2em\hbox{\rm R}}}}
\def\IH{{\hbox{{\rm I}\kern-.2em\hbox{\rm H}}}}
\def\IC{{\ \hbox{{\rm I}\kern-.6em\hbox{\bf C}}}}
\def\IZ{{\hbox{{\rm Z}\kern-.4em\hbox{\rm Z}}}}

\newcommand{\beq}{\begin{equation}}
\newcommand{\eeq}{\end{equation}}

\newcommand{\CQG}[1]{{\sl Class.~Quant.\ Grav.} {\bf #1}}
\newcommand{\PRD}[1]{{\sl Phys.~Rev.} {\bf D#1}}

\newcommand{\JMP}[1]{{\sl J.~Math.~Phys.} {\bf #1}}
\begin{document}
%
%
%
%
\def\citen#1{%
\edef\@tempa{\@ignspaftercomma,#1, \@end, }
\edef\@tempa{\expandafter\@ignendcommas\@tempa\@end}%
\if@filesw \immediate \write \@auxout {\string \citation
{\@tempa}}\fi
\@tempcntb\m@ne \let\@h@ld\relax \let\@citea\@empty
\@for \@citeb:=\@tempa\do {\@cmpresscites}%
\@h@ld}
%
\def\@ignspaftercomma#1, {\ifx\@end#1\@empty\else
   #1,\expandafter\@ignspaftercomma\fi}
\def\@ignendcommas,#1,\@end{#1}
%
%
\def\@cmpresscites{%
 \expandafter\let \expandafter\@B@citeB \csname b@\@citeb
\endcsname
 \ifx\@B@citeB\relax 
    \@h@ld\@citea\@tempcntb\m@ne{\bf ?}%
    \@warning {Citation `\@citeb ' on page \thepage \space
undefined}%
 \else
    \@tempcnta\@tempcntb \advance\@tempcnta\@ne
    \setbox\z@\hbox\bgroup 
    \ifnum\z@<0\@B@citeB \relax
       \egroup \@tempcntb\@B@citeB \relax
       \else \egroup \@tempcntb\m@ne \fi
    \ifnum\@tempcnta=\@tempcntb 
       \ifx\@h@ld\relax 
          \edef \@h@ld{\@citea\@B@citeB}%
       \else 
          \edef\@h@ld{\hbox{--}\penalty\@highpenalty \@B@citeB}%
       \fi
    \else   
       \@h@ld \@citea \@B@citeB \let\@h@ld\relax
 \fi\fi%
 \let\@citea\@citepunct
}
%
\def\@citepunct{,\penalty\@highpenalty\hskip.13em plus.1em
minus.1em}%
%
%
\def\@citex[#1]#2{\@cite{\citen{#2}}{#1}}%
%
%
\def\@cite#1#2{\leavevmode\unskip
  \ifnum\lastpenalty=\z@ \penalty\@highpenalty \fi 
  \ [{\multiply\@highpenalty 3 #1
      \if@tempswa,\penalty\@highpenalty\ #2\fi 
    }]\spacefactor\@m}
\let\nocitecount\relax  
%
\begin{titlepage}
\vspace{.5in}
\begin{flushright}
\vspace{.25in}
UCD-PHY-95-40\\
gr-qc/9511060\\
November 1995\\
\end{flushright}
\vspace{.5in}
\begin{center}
{\Large\bf
The symplectic structure of general relativity\\in the double-null
(2+2) formalism
}\\
\vspace{.4in}
{R{\sc ichard}~J.~E{\sc pp}\footnote{\it email:
repp@dirac.ucdavis.edu}\\
       {\small\it Department of Physics}\\
       {\small\it University of California}\\
       {\small\it Davis, CA 95616}\\{\small\it USA}}
\end{center}

\vspace{.5in}
\begin{center}
\begin{minipage}{6in}
\begin{center}
{\large\bf Abstract}
\end{center}
{\small
In the (2+2) formulation of general relativity spacetime is
foliated by a
two-parameter family of spacelike 2-surfaces (instead of the more
usual
one-parameter family of spacelike 3-surfaces).  In a partially
gauge-fixed setting (double-null gauge), I write down the
symplectic
structure of general relativity in terms of intrinsic and extrinsic
quantities associated with these 2-surfaces.  This leads to an
identification of the reduced phase space degrees of freedom.  In
particular, I show that the two physical degrees of freedom of
general
relativity are naturally encoded in a quantity closely related to
the twist of
the pair of null normals to the 2-surfaces.  By considering the
characteristic initial-value problem I establish a canonical
transformation
between these and the more usually quoted conformal 2-metric
(or shear)
degrees
of freedom.
(This paper is based on a talk given at the Fifth Midwest
Relativity
Conference, Milwaukee, USA.)
}
\end{minipage}
\end{center}
\end{titlepage}
\addtocounter{footnote}{-1}

The (2+2) formulation of general relativity was first introduced by
Sachs \cite{Sachs}, and has since then attracted the interest of
many researchers \cite
{Geroch,d'Inverno,Smallwood,Torre,Hayward,Yoon,d'Inverno2,Brady}.
The basic idea is to foliate spacetime by a
two parameter family of spacelike 2-surfaces, rather than the more
usual one parameter family of spacelike 3-surfaces.  One can then
construct 3-surfaces by stringing together one parameter families
of these spatial 2-surfaces.  Since the geometry of spacetime is
encoded in intrinsic and extrinsic quantities associated with the
spatial 2-surfaces only, the 3-surfaces can just as easily be
spacelike, timelike, or null.  In particular, one can avoid having
to deal with the degenerate 3-metric on a null 3-surface.

The formalism is ideally suited to the study of spacetimes
(or subsets thereof)
with topology $R^{2}\times S$, where $S$ is a spacelike
compact 2-surface
without boundary.  A good example is the extended Schwarzschild
black hole.  Furthermore, in the double-null gauge of the (2+2)
formalism spacetime is foliated by two sets of null 3-surfaces,
which intersect in the spacelike 2-surfaces $S$.
The bifurcate Killing horizon of the Schwarzschild black hole,
for instance, can be identified with a pair of such null
3-surfaces, intersecting in the bifurcation 2-sphere.
Also, a quantity such as the expansion of null geodesics
normal to $S$ (used
in the definitions of an apparent horizon
and a trapped surface, for example \cite{Wald}) is a natural
quantity in the (2+2) formalism.
In short, this formalism provides a very convenient setting for the
study of many questions in general relativity.
For a recent, more comprehensive introduction see
\cite{Brady}.

The main new result presented here
is the expression of the symplectic structure of general relativity
in terms of intrinsic and extrinsic quantities associated with the
spatial 2-surfaces of the double-null (2+2) formalism.
I also introduce a new variable, closely
related to the twist of the null normals to the 2-surfaces, which
encodes the two physical gravitational degrees of freedom.  This
provides an interesting alternative to the more usual
conformal 2-metric (or shear)
degrees of freedom.

The paper is organized as follows.  I begin with an introduction
to the ideas and language of the double-null (2+2) formalism, in
particular the intrinsic and extrinsic quantities associated with
the spatial 2-surfaces.  Then I say a few words about gauge-fixing
in order to understand geometrically how the twist encodes
gravitational degrees of freedom.  The symplectic structure is then
presented, in two versions: one emphasizing the twist degrees of
freedom, and the other the shear (or conformal 2-metric) degrees of
freedom.  In passing, I make a few comments about what is held
fixed in the action principle when the spacetime boundary is null
(at least for the Einstein-Hilbert action).  Finally, I discuss
the characteristic initial-value problem.  Since the phase space
can be identified with the space of initial data, this helps to
illuminate the symplectic structure results.  It also allows one to
establish a canonical transformation between the twist and shear
sets of degrees of freedom.

\section{Double-null (2+2) formalism}

We consider a manifold $M=R^{2}\times S$, where $S$ is a
compact 2-surface without boundary.  For simplicity we restrict
ourselves
to the case $S=S^{2}$, which is relevant for black hole
spacetimes.\footnote{However, surfaces of higher genus may
introduce
interesting topological degrees of freedom into the following
analysis \cite{Epp2}.}
Introduce
coordinates $x^{A}$ ($A,B,\ldots=+,-$) for $R^{2}$ and (local)
coordinates
$x^{i}$ ($i=1,2$) for the 2-sphere.  Now equip $M$ with a
(time-orientable) Lorentzian metric
and apply a suitable (active) diffeomorphism such that the
3-surfaces
$x^{\pm}=const$ (denoted by $\Sigma^{\mp}$) are null, and their
intersecting
2-surfaces $S$ are spacelike.
We shall work in this so-called double-null gauge, in which the
spacetime
metric takes the form
\beq
g_{ab}=h_{ab}-2e^{-\lambda}\partial_{(a}x^{+}\partial_{b)}x^{-},
\eeq
where $h_{ab}$ is the induced metric on $S$,
and $a,b,\ldots$ denote abstract spacetime indices.
At any point $p\in M$
the subspace of $T_{p}M$
orthogonal to $T_{p}S$ is spanned by null vector
fields
of the form
\beq
n_{A}^{a}:=(\partial_{A})^{a}-s_{A}^{a},
\label{null normals}
\eeq
where
$\partial_{A}:=\partial/\partial x^{A}$, and $s_{A}^{a}$ are shift
vectors
lying in $T_{p}S$.  See Figure 1.
It is assumed that both $n_{A}^{a}$ are future-pointing, and since
\beq
g_{ab}n_{+}^{a}n_{-}^{b}=-e^{-\lambda},
\label{normalization}
\eeq
we see that the scalar $\lambda$ is associated with their
normalization.

Now we introduce the extrinsic fields, which measure how $S$ is
imbedded into $M$.
First there is the extrinsic curvature, which carries an
index $A$ since there are now two normal directions to
consider:
\beq
K_{Aab}:=\perp {1\over2}{\cal L}_{A}h_{ab}.
\eeq
Here ${\cal L}_{A}$ denotes the Lie derivative with respect to
$n_{A}^{a}$, and $\perp$
means spatial projection of all indices to the right by
$h^{a}_{\;\;b}$.
The trace of the extrinsic curvature is called the expansion:
\beq
\theta_{A}:=h^{ab}K_{Aab},
\eeq
whose physical interpretation is clear from the relation
\beq
\perp{\cal L}_{A}\epsilon=\epsilon\theta_{A},
\eeq
where $\epsilon$ denotes the volume form on $S$.
In this relation $\perp$ acts on the volume form indices,
which
for convenience are suppressed.
The trace-free part,
\beq
\sigma_{Aab}:=K_{Aab}-{1\over2}\theta_{A}h_{ab},
\eeq
is called the shear.
Defining the (inverse) conformal 2-metric
\beq
\tilde{h}^{ab}:=\epsilon h^{ab},
\eeq
its Lie derivative along the null normals is the shear density:
\beq
\tilde{\sigma}_{A}^{ab}:=\epsilon\sigma_{A}^{ab}=
-\perp {1\over2}{\cal L}_{A}\tilde{h}^{ab}.
\label{shear density}
\eeq
In mixed indices the shear has the standard form of a current,
bilinear in the conformal 2-metric:
\beq
\sigma_{A\;\;c}^{\;\;a}=
\perp {1\over4}\tilde{h}^{ab}{\cal L}_{A}\tilde{h}_{bc},
\eeq
and can be interpreted as a gravitational wave current (see, for
example, \cite{Yoon}).  Here $\tilde{h}_{ab}:=\epsilon^{-1}h_{ab}$
is the conformal 2-metric.

Another extrinsic quantity, one which will play an important role
in our discussions, is the [normalized---cf (\ref{normalization})]
twist
\beq
\omega^{a}:=-e^{\lambda}[n_{+},n_{-}]^{a},
\eeq
which is tangent to $S$ and measures the nonintegrability of the
null normals.
Finally, there is the ``inaffinity" \cite{Hayward}
\beq
\nu_{A}:={\cal L}_{A}\lambda.
\eeq
Its name derives from the fact that
\beq
n_{A}^{b}\nabla_{b}n_{A}^{a}=-\nu_{A}n_{A}^{a}\;\;\;
({\rm no}\;{\rm sum}\;{\rm on}\;A),
\eeq
so the
$n_{A}^{a}$ generate
null geodesics, but generally with non-affine parametrization.
(Here $\nabla$ denotes the
spacetime covariant derivative operator.)

\section{Gauge-fixing and true degrees of freedom}

It is well known that the gravitational field in general relativity
has two degrees of freedom per space point.  In the double-null
(2+2) formalism there are several quantities that have two
independent components, each of which is therefore a good candidate
for encoding the two gravitational degrees of freedom: the shear,
being symmetric and trace-free; the conformal 2-metric, which has
unit determinant (in the dyad basis); and the twist, which is a
vector tangent to $S$.  Of these three, the first two are closely
related
[see (\ref{shear density})], and usually it is one or the other of
these which is used to represent the true gravitational degrees of
freedom.  In this paper I will emphasize instead the twist,
so it
is instructive at this point to say a few words about gauge-fixing,
and elaborate on the physical interpretation of the twist.

Let us restrict our attention to the wedge of spacetime defined by
$x^{A}\geq 0$.  The null 3-surfaces $x^{\pm}=0$ which bound this
region are denoted by $\Sigma^{\mp}_{0}$, and their intersection by
$S_{0}$.  The $\Sigma^{\pm}$  3-surfaces should be thought of as
congruences of null geodesics, as shown in Figure 2a.  Now consider
an (active) diffeomorphism generated by the vector field
\beq
\xi^{a}:=\xi_{\perp}^{a}+\xi_{\parallel}^{a}:=
\xi_{\perp}^{i}(\partial_{i})^{a}+
\xi_{\parallel}^{A}n_{A}^{a},
\eeq
where $\partial_{i}:=\partial/\partial x^{i}$.  It turns out that
the double-null gauge is preserved provided the $\xi_{\parallel}$
diffeomorphisms satisfy the restrictions
\beq
{\cal L}_{\pm}\xi_{\parallel}^{\mp}=0.
\label{DN gauge}
\eeq
Geometrically, such $\xi_{\parallel}$ diffeomorphisms correspond to
moving null geodesics from any one $\Sigma^{A}$ plane to other
$\Sigma^{A}$ planes, such that all these planes remain congruences
of null geodesics.
A gauge-fixing condition we shall find useful, and which is always
reachable, is
\beq
\mu_{A}:=\nu_{A}+\kappa\theta_{A}=0
\;\;\;{\rm on}\;\;\;\Sigma^{A}_{0},
\label{mu gauge fix}
\eeq
where $\kappa$ is any constant (to be chosen later).
Although this does not completely fix the $\xi_{\parallel}$
gauge\footnote{A further condition might be to restrict
$\xi^{\pm}=0$
on $S_{0}$ [and hence on $\Sigma^{\mp}_{0}$---see
(\ref{DN gauge})], i.e. to not allow the
movement of null geodesics into or out of the wedge $x^{A}\geq 0$.
But even in this case there remains a residual set of nontrivial
$\xi_{\parallel}$
diffeomorphisms which preserves (\ref{mu gauge fix}) \cite{Epp1}.}
it is sufficient for our present purposes.

Turning attention to the remaining ``$S$-diffeomorphisms"---those
acting along the ``fibers" $S$---it is clear from Figure 2a that
one can always do a series of such $\xi_{\perp}$ diffeomorphisms to
``straighten-out" the null geodesics in, say, all of the
$\Sigma^{+}$ planes.  This
corresponds to gauge-fixing one of the shift vectors to zero:
\beq
s_{+}^{a}=0.
\eeq
Furthermore, one can still do $S$-diffeomorphisms which are
the same for each $S$ foliating a given $\Sigma^{+}$ (i.e.
independent of $x^{+}$),
but which may differ from one $\Sigma^{+}$ to another.  This
freedom can be exactly used up by straightening-out the null
geodesics on just one $\Sigma^{-}$, say $\Sigma^{-}_{0}$.  This
corresponds to the gauge-fixing condition
\beq
s_{-}^{a}=0\;\;\;{\rm on}\;\;\;\Sigma^{-}_{0}.
\label{s- gauge fix}
\eeq
Now, while this does not completely fix the $\xi_{\perp}$
gauge\footnote{One is still free to do $S$-diffeomorphisms which
are the same for each $S$ foliating the spacetime wedge $x^{A}\geq
0$ (i.e. independent of $x^{+}$ and $x^{-}$).  Up to a conformal
Killing vector subtlety, this freedom
can be exactly used up by fixing the inverse conformal 2-metric
$\tilde{h}^{ab}$ on $S_{0}$ \cite{Epp1}.} it is again sufficient
for our
purposes to stop here.  See Figure 2b.
Note that in this
hierarchy of gauge-fixing conditions
[(\ref{mu gauge fix})--(\ref{s- gauge fix})], reaching each
successive
condition preserves the previous ones.

The important point is that (almost) all of the gauge degrees of
freedom have now been fixed, so any nontrivial degrees of freedom
remaining must be physical: these are the two components of
the twist, which encode, for example, the failure of points $p$ and
$p^{'}$ in Figure 2b to coincide.  In fact, inspection of
the symplectic structure (discussed in the next section) reveals
that the relevant quantity is actually a slightly modified form of
the twist:
\beq
\omega_{\pm a}:=\pm\omega_{a}+D_{a}\lambda,
\eeq
where $D$ is the covariant derivative operator in $S$.  We remark,
however, that this cannot be the whole story:  the Schwarzschild
solution, for example, has $\omega_{\pm a}=0$ for all values of the
mass.  This subtlety is related to the word ``almost" used in
parenthesis at the beginning of this paragraph \cite{Epp1}.  The
Kerr example will be analysed in
detail elsewhere, where it is expected that the twist will be
proportional to the black hole angular momentum \cite{Epp2}.

\section{Action principle and symplectic structure}

It has long been known that the phase space of a classical system
can be understood in a covariant way (i.e. with no preferred time
slice) by considering it to be the space of classical solutions,
${\cal S}$ \cite{Ashtekar}.
We shall briefly review the standard construction
of the symplectic structure on ${\cal S}$
(see, for example, \cite{Burnette}), adapted here
to the (2+2) splitting of spacetime.

Let us consider the quantity of action in an ``evolution region"
${\cal E}$ of $M$, defined by $0\leq x^{A}\leq 1$ (see Figure
3):\footnote{Note that there is no loss of generality in taking
endpoints at $x^{A}=1$ since we can always effectively rescale the
$x^{A}$-axes by (active) diffeomorphisms of the fields.}
\beq
I_{\cal E}=\int_{\cal E}\,dx^{+}\,dx^{-}\,L(\varphi).
\eeq
The Lagrangian $L(\varphi)$ is a functional of fields, collectively
denoted as $\varphi$, and takes the form of an integral over $S$.
Variation of the action results in a term proportional to the
Euler-Lagrange equations, which vanishes when $\varphi\in {\cal
S}$, leaving the surface term:
\beq
\delta I_{\cal E}=\int_{\partial {\cal E}}
\left\{J^{+}(\varphi,\delta\varphi)\,dx^{-}-
J^{-}(\varphi,\delta\varphi)\,dx^{+}\right\}.
\label{delta I}
\eeq
This defines the current components
$J^{A}(\varphi,\delta\varphi)$, at least up to the ambiguity
\beq
J^{\pm}(\varphi,\delta\varphi)\mapsto
J^{\pm}(\varphi,\delta\varphi)\pm\partial_{\mp}
Z(\varphi,\delta\varphi),
\label{J ambiguity}
\eeq
for arbitrary
$Z=Z(\varphi,\delta\varphi)$.
This ambiguity will be exploited in section~3.2 below.

Now consider any 3-surface $\Sigma$ consisting of a one parameter
family of spatial 2-surfaces $S$ stretching between $S_{L}$ and
$S_{R}$, as shown in Figure 3.
The presymplectic potential is defined as
\beq
\Theta_{\Sigma}:=\int_{\Sigma}
\left\{J^{+}(\varphi,\delta\varphi)\,dx^{-}-
J^{-}(\varphi,\delta\varphi)\,dx^{+}\right\},
\label{Theta}
\eeq
and the presymplectic structure as its second antisymmetrized
variation:
\beq
\Omega=\int_{\Sigma}
\left\{\Omega^{+}(\varphi,\delta_{1}\varphi,
\delta_{2}\varphi)\,dx^{-}-
\Omega^{-}(\varphi,\delta_{1}\varphi,\delta_{2}\varphi)\,dx^{+}
\right\},
\eeq
where
\beq
\Omega^{A}(\varphi,\delta_{1}\varphi,\delta_{2}\varphi):=
\delta_{1}J^{A}(\varphi,\delta_{2}\varphi)-(1\leftrightarrow 2).
\label{Omega A}
\eeq
Here $\delta_{\mu}\varphi\in T_{\varphi}{\cal S}$, $\mu=1,2$, and
can be thought of as the partial derivative of $\varphi$ with
respect to some (suppressed) solution space coordinates.
$\Omega$ is the presymplectic\footnote{Presymplectic (as opposed to
symplectic) refers to the fact that $\Omega$ has degenerate
directions on $T{\cal S}$; these are tangent to
the gauge orbits of the
theory \cite{Ashtekar}.  An example will be given in section~3.2
below.}
structure evaluated at the phase space point
$\varphi\in{\cal S}$, contracted with the two ``vector fields"
$\delta_{1}\varphi$ and $\delta_{2}\varphi$, which are solutions to
the linearized Euler-Lagrange equations.

It is easy to show that, while $\Theta_{\Sigma}$ in general depends
on the choice of $\Sigma$ interpolating between $S_{L}$ and
$S_{R}$, $\Omega$ does not.  For instance, we may equally well use
either the Cauchy surface $\Sigma$ or the characteristic surface
$\Sigma^{'}$ in Figure 3 to evaluate $\Omega$.  This is what is
meant by the covariance of the phase space.  It should also be
noted that $\Omega$ is invariant under the addition of boundary
terms to the action.

\subsection{Action principle}

For the Einstein-Hilbert action we obtain
\beq
J^{\pm}(\varphi,\delta\varphi)=
\int_{S}\,\epsilon\,\left\{-\sigma_{\mp ab}(\delta h)_{T}^{ab}
-\omega_{\mp a}\delta s_{\mp}^{a}+\delta(\theta_{\mp}-2\nu_{\mp})
\right\}
+\partial_{\mp}\int_{S}\,\epsilon\,\delta(\ln\sqrt{h}+\lambda).
\label{J sigma form}
\eeq
First observe that $J^{\pm}$ consists of a ``bulk" term and an
``edge" term\footnote{The distinction between bulk and edge terms
is not unambiguous; the form written down here is based partly on
experience working with the equations.  Note in particular that the
combination $(\theta_{\pm}-2\nu_{\pm})$ appears naturally also in
the Euler-Lagrange equation (\ref{shift eq}).}---the latter makes
contributions to
(\ref{delta I}) only at the ``edges" $S_{0}$,
$S_{1}$, $S_{L}$, and $S_{R}$ (see Figure 3).  Next,
\beq
\delta h^{ab}=-h^{ab}\delta\ln\sqrt{h}+(\delta h)_{T}^{ab}
\label{delta h}
\eeq
has been split into its trace and trace-free parts,
respectively, where $h$ is the determinant of the spatial 2-metric
in the dyad basis.
Following the
standard decomposition of a trace-free symmetric tensor
\cite{York1}
we can write
\beq
(\delta h)_{T}^{ab}=(\delta h)_{TT}^{ab}+(Lv)^{ab}.
\eeq
Here $(\delta h)_{TT}^{ab}$ is the transverse trace-free part of
$\delta h^{ab}$, i.e. it is trace-free and satisfies
$D_{a}(\delta h)_{TT}^{ab}=0$, where, as noted above, $D$ is the
covariant derivative operator in $S$.
In general, the transverse trace-free sector is spanned by
the Teichm\"{u}ller parameters associated with the topology of $S$
\cite{Mazur}, but for our choice ($S=S^{2}$) this space is empty.
All that remains is the ``exact" part:
\beq
(Lv)^{ab}:=D^{a}v^{b}+D^{b}v^{a}-h^{ab}D_{c}v^{c},
\eeq
which defines the vector $v^{a}\in TS$ uniquely up to conformal
Killing vectors (CKV)---if any exist.

The form of $J^{A}$ given in (\ref{J sigma form}) has the conformal
metric and shear playing a dominant role.  However, in this paper
we wish to emphasize the twist degrees of freedom.
To achieve this we integrate the $(Lv)^{ab}$ term by parts and use
the following Euler-Lagrange equation:
\beq
\perp{\cal L}_{\pm}\tilde{\omega}_{\pm a}=
\epsilon 2D^{b}\sigma_{\pm ab}-
\epsilon D_{a}(\theta_{\pm}-2\nu_{\pm}),
\label{shift eq}
\eeq
where $\tilde{\omega}_{\pm a}:=\epsilon{\omega}_{\pm a}$.  After
also integrating $\perp{\cal L}_{\pm}\tilde{\omega}_{\pm a}$ by
parts we get
\beq
J^{\pm}(\varphi,\delta\varphi)=
\int_{S}\,\left\{\tilde{\omega}_{\mp a}\,\Delta n_{\mp}^{a}
+\epsilon\,\Delta(\theta_{\mp}-2\nu_{\mp})\right\}
+\partial_{\mp}\int_{S}\,\left\{\tilde{\omega}_{\mp a}v^{a}
+\epsilon\,\delta(\lambda+\ln\sqrt{h})\right\}.
\label{J omega form}
\eeq
Here we used $\delta s_{A}^{a}=-\delta
n_{A}^{a}$, which is obvious from (\ref{null normals}).  Note that
\beq
\Delta:=\delta+{\cal L}_{v},
\eeq
is a {\em manifestly $S$-diffeomorphism invariant} variation.
This is because under an $S$-diffeomorphism generated by
$\xi_{\perp}^{a}$ we have $\delta\varphi=-{\cal
L}_{\xi_{\perp}}\varphi$  (at least for the fields $\varphi$ that
$\Delta$ will be applied to---for example, it is not true for the
shift vectors).  On the other hand, it turns out that
$v^{a}=\xi_{\perp}^{a}$ (up to CKV), so $\Delta\varphi=0$ (up to a
possible CKV term).  Note that the ambiguity of $v^{a}$ up to CKV
appears to be a problem
because $v^{a}$---not just $(Lv)^{ab}$---appears in the above
formula for $J^{A}$.  But in fact, $J^{A}$
(bulk {\em plus} edge terms) is invariant under $v^{a}\mapsto
v^{a}+v^{a}_{\rm CKV}$, which is actually obvious since the
original form
[see (\ref{J sigma form})] depends on only $(Lv)^{ab}$.
So in the case of $S$-diffeomorphisms the only contribution to
$J^{A}$ comes from the edge terms, the result being
\beq
J^{\pm}(\varphi,\delta\varphi)=
\mp\partial_{\mp}\int_{S}\tilde{\omega}_{a}\xi_{\perp}^{a}.
\label{exact}
\eeq
Notice that this produces an exact differential in the integrand of
(\ref{delta I}), and so, since $\partial\partial{\cal E}\equiv 0$,
the action is indeed invariant under $S$-diffeomorphisms, as
it should be.

The answer to the question of what is fixed on the boundary in the
action principle of general relativity is well understood, at least
for boundaries composed of spacelike and timelike sections
\cite{York2}.  The double-null (2+2) fomalism appears to be ideally
suited to extending our understanding to the case of null
boundaries.  Let us make one comment in this regard.
Ignoring all but the ``$\tilde{\omega}\Delta n$" term in
(\ref{J omega form}) we
learn that {\em what is fixed on a null boundary are the generators
of the null geodesics, up to $S$-diffeomorphisms}.  This
represents two ``$q$"s---half of the reduced phase space degrees of
freedom.  This simple answer is complicated by the other bulk term
(which is associated with the parametrization of the null
geodesics), as well as the edge terms.  Also, of course, the
question of what is held fixed is sensitive to boundary terms added
to the Einstein-Hilbert action (to make it first order, for
instance)
\cite{Epp1}.

It is instructive at this point to draw an analogy with York's
classic results \cite{York2} on the
conformal 3-metric degrees of freedom of the gravitational field.
For spacetime $M=R\times \Sigma$, $\Sigma$ spacelike and closed,
the variation of the ``cosmological action" $S_{K}$ is
\beq
\delta S_{K}=\int_{\Sigma_{2}-\Sigma_{1}}\,d^{3}x\,
(\tilde{p}^{ij}\delta\tilde{\gamma}_{ij}+p_{K}\delta K).
\eeq
Like the ``$\tilde{\omega}\Delta n$" term in
(\ref{J omega form}), the ``$\tilde{p}\delta\tilde{\gamma}$" term
represents two independent ``$q$"s: five in the conformal 3-metric
$\tilde{\gamma}_{ij}$, minus three because $\delta S_{K}=0$ for
spatial diffeomorphisms.  The other bulk term, $p_{K}\delta K$,
where $K$ is the trace of the extrinsic curvature
of $\Sigma$, is
``trivialized" by the {\em gauge choice} $K=const$ on the spatial
3-surfaces.  The (2+2) analogue of $K$ appears to be
$(\theta-2\nu)$, and the corresponding gauge choice would be
$(\theta_{\mp}-2\nu_{\mp})=const$ (or zero) on the
spatial 2-surfaces foliating $\Sigma^{\mp}_{0}$.  This gauge choice
is discussed more in the following subsection.

\subsection{Symplectic structure}

Finally, let us calculate the symplectic structure.  Equation
(\ref{Omega A}) instructs us to take the second variation of
$J^{A}$ and antisymmetrize.  Using the fact that partials commute
(so $\delta_{1}\delta_{2}\varphi=\delta_{2}\delta_{1}\varphi$), the
only further bit of information we need is
\beq
\delta_{2}v_{1}^{a}-\delta_{1}v_{2}^{a}=[v_{1},v_{2}]^{a},
\eeq
which follows from the fact that the second antisymmetrized
variation of both sides of (\ref{delta h}) must vanish.  We find
\begin{eqnarray}
\Omega^{\pm}(\varphi,\delta_{1}\varphi,\delta_{2}\varphi)&=&
\int_{S}\,\left\{\Delta_{1}\tilde{\omega}_{\mp a}\,\Delta_{2}
n_{\mp}^{a}
+\Delta_{1}\epsilon\,\Delta_{2}(\theta_{\mp}-2\nu_{\mp})\right\}
\label{Omega omega form}\\
&&+\partial_{\mp}\int_{S}\,\left\{\Delta_{1}
\tilde{\omega}_{\mp a}v_{2}^{a}
+{1\over2}\tilde{\omega}_{\mp a}[v_{1},v_{2}]^{a}
+\delta_{1}\epsilon\,\delta_{2}\lambda\right\}-(1\leftrightarrow
2).
\nonumber
\end{eqnarray}
The bulk term has the standard $\Delta p\wedge\Delta q$
form, which
allows us to immediately identify
canonically conjugate pairs of phase space variables.  Note the
appearance again of the $S$-diffeomorphism invariant variation,
$\Delta$.

The edge term in (\ref{Omega omega form}) can be simplified as
follows.  As pointed out earlier, the choice of
$J^{\pm}(\varphi,\delta\varphi)$ has the arbitrariness displayed in
equation
(\ref{J ambiguity}).  Equation
(\ref{exact}) suggests we chose
\beq
Z(\varphi,\delta\varphi)=\int_{S}\,\tilde{\omega}_{a}v^{a},
\label{chose Z}
\eeq
in which case it can be shown that
\beq
\Omega^{\pm}(\varphi,\delta_{1}\varphi,\delta_{2}\varphi)
\longmapsto\ldots +\partial_{\mp}\int_{S}\,
\Delta_{1}\epsilon\,\Delta_{2}\lambda-(1\leftrightarrow 2).
\label{new omega}
\eeq
Thus, in this form at least, $S$-diffeomorphisms are {\em
manifestly} associated with degenerate directions of the symplectic
structure, i.e. they are
gauge transformations.  This is not surprising, of course, but on
the other hand, the situation is not nearly so clear for the
$\xi_{\parallel}$ diffeomorphisms [satisfying
(\ref{DN gauge})].  These will be analyzed elsewhere \cite{Epp1},
and are expected to be associated with so-called edge degrees of
freedom.  (For a recent review of edge degrees of freedom and their
possible role in explaining black hole entropy, see \cite{Carlip}.)

Let us now remark on the repeated appearance of the combination of
variables $(\theta_{\pm}-2\nu_{\pm})$ throughout the foregoing
analysis.  Recalling the gauge-fixing condition
(\ref{mu gauge fix}), we see that {\em the symplectic structure
provides strong support for the choice} $\kappa=-1/2$,
which would then eliminate an ``unwanted" bulk term in
(\ref{Omega omega form}).  Comparing with the literature (for
example \cite{Sachs}) we find
instead that the usual choice is $\kappa=0$:
affine
parametrization of the null geodesics on $\Sigma^{A}_{0}$.
Alternatively, Hayward \cite{Hayward} points out that the choice
$\kappa=+1/2$ simplifies (linearizes) one of the Euler-Lagrange
equations, namely the Raychaudhuri, or ``focusing", equation.  But
this is {\em opposite} in sign to the $\kappa$ required to simplify
the symplectic structure!  It appears that there is some
fundamental complexity here which shows up in one place or the
another---it cannot be gauge-fixed away.

For completeness let us record also the ``shear form" of the
symplectic structure, obtained straightforwardly from
(\ref{J sigma form}), and the fact that
$\epsilon(\delta h)_{T}^{ab}=\perp\delta\tilde{h}^{ab}$:
\begin{eqnarray}
\Omega^{\pm}(\varphi,\delta_{1}\varphi,\delta_{2}\varphi)&=&
\int_{S}\left\{-\delta_{1}\sigma_{\mp ab}\,
\delta_{2}\tilde{h}^{ab}-\delta_{1}\tilde{\omega}_{\mp a}\,
\delta_{2}s_{\mp}^{a}+\delta_{1}\epsilon\,
\delta_{2}(\theta_{\mp}-2\nu_{\mp})\right\}
\nonumber\\
&&+\partial_{\mp}\int_{S}\delta_{1}\epsilon\,\delta_{2}\lambda
-(1\leftrightarrow 2).
\label{Omega sigma form}
\end{eqnarray}
If one imposes the gauge-fixing conditions discussed in section~2
[equations (\ref{mu gauge fix}-\ref{s- gauge fix})] the only bulk
term that survives (on $\Sigma^{A}_{0}$ at least) is the shear
term.  Like the twist term in
(\ref{Omega omega form}) it also encodes the two true degrees of
freedom of the theory.  However, ignoring for a moment the
$(\theta-2\nu)$ term common to both descriptions, the twist term
does so {\em covariantly}---without having to gauge-fix the shift
vectors.  The advantage is that the twist form is thus applicable
even when $\Sigma$ is not a characteristic 3-surface, in particular
it can be taken to be Cauchy (see Figure 3).  This may have
important implications in a quantization programme.
In this sense the (modified) twist is a more natural choice than
the shear to describe the true degrees of freedom.

To help better understand both forms of the symplectic structure it
is instructive to now examine the characteristic initial-value
problem, since the phase space can also be thought of as the set of
initial data.

\section{Characteristic initial-value problem}

The earliest dicussion of the characteristic initial-value problem
for general relativity is due to Sachs \cite{Sachs}, followed by
others.
We shall not repeat this analysis here, but merely
quote the main result:
the initial data required to
obtain a unique solution in the wedge $x^{A}\geq 0$ is indicated in
Figure 4a.\footnote{
Note that we do not gauge-fix any of the initial data; we shall elaborate
elsewhere on the splitting of this data into physical and gauge parts
\cite{Epp1}.
Also, we are not concerned here with the question of
existence of solutions, or caustics they may develop.}
The placement of the various fields in the figure has the following
meaning: $s_{+}^{a}$ is specified everywhere in the wedge;
$s_{-}^{a}$ is specified only on $\Sigma^{-}_{0}$; $\tilde{h}^{ab}$
is specified only on $S_{0}$; etc.
This might be called the ``shear version" of the initial-value
problem, where $\tilde{\sigma}_{\pm}^{ab}$ on $\Sigma^{\pm}_{0}$ is
the main physical data.
In Figure 4b
we introduce a ``twist version", which emphasizes the alternative
data
$\tilde{\omega}_{\pm a}$ on $\Sigma^{\pm}_{0}$.
It is instructive to at least outline the proof that these two sets
of initial data are equivalent.

Start with the data in Figure 4b, which provides, in particular,
$\perp{\cal L}_{\pm}\tilde{\omega}_{\pm a}$ and
$(\theta_{\pm}-2\nu_{\pm})$ on $\Sigma^{\pm}_{0}$.  However, we
know $\sqrt{h}$---and hence $\epsilon$---only on $S_{0}$.  Thus,
the Euler-Lagrange equation
(\ref{shift eq}) gives us $D^{b}\sigma_{\pm ab}$ on $S_{0}$ (only).
Now, using the fact that $\sigma_{\pm ab}$ is trace-free
(and $S=S^{2}$) we can
write
\beq
\sigma_{\pm ab}=(L\tau_{\pm})_{ab}:=D_{a}\tau_{\pm b}+
D_{b}\tau_{\pm a}-h_{ab}D^{c}\tau_{\pm c},
\eeq
which defines $\tau_{\pm}^{a}\in TS$ uniquely up to CKV.  (Note
that we know the full spatial metric on $S_{0}$.)  Thus,
\beq
D^{b}\sigma_{\pm ab}=(D^{2}+{1\over2}R)\tau_{\pm a},
\label{canonical transformation}
\eeq
where $D^{2}$ is the Laplacian and $R$ the Ricci scalar on $S$.
Given the left hand side,\footnote{Notice that in order for
solutions to exist the left hand side, or source term, must be
orthogonal to the kernel of $(D^{2}+{1\over2}R)$ \cite{York1}.
This
amounts to restrictions on the twist initial data such that, for
example, $J^{A}$ is invariant under $v^{a}\mapsto
v^{a}+v^{a}_{\rm CKV}$, as discussed in the paragraph containing
equation
(\ref{J omega form}) \cite{Epp1}.} this
equation can be solved for $\tau_{\pm}^{a}$, unique up to CKV
\cite{York1}.  We can then {\em uniquely} determine
$\tilde{\sigma}_{\pm}^{ab}$ on $S_{0}$.  From
here we use various Euler-Lagrange equations to iterate our
way up $\Sigma^{\pm}_{0}$, at each step using the {\em nonlocal}
transformation
(\ref{canonical transformation})---the heart of the canonical
transformation from Figure 4b to 4a initial data.  It is easy to
show also the converse, from Figure 4a to 4b.

Finally, let us compare the initial data in Figures 4a and 4b with
the corresponding symplectic structures:
(\ref{Omega sigma form}) and (\ref{Omega omega form}).  The latter
have the general form $\delta p\wedge\delta q$.  Now normally $p$
and $q$ are independent, and can be identified with the initial
data, but here the {\em characteristic} initial data tells us that
only $p$ is independent, and $q$ is to be determined by integrating
the Euler-Lagrange equations up $\Sigma^{\pm}_{0}$ from $S_{0}$.
This is because essentially $p=\partial_{\pm}q$ on
$\Sigma^{\pm}_{0}$; the independent data can be thought of as $p$
on $\Sigma^{\pm}_{0}$ and $q$ on $S_{0}$.  In the symplectic
structure, $q$ on $\Sigma^{\pm}_{0}$ is then determined
{\em nonlocally}
from this independent data by integration.  Only when $\Sigma$ is
a Cauchy surface
do we recover the usual local symplectic structure, with $p$ and
$q$ independent, and so on.

\section{Concluding remarks}

The (2+2) formulation of general relativity has resurfaced
time and again since 1962 \cite{Sachs,Geroch,d'Inverno,Smallwood,
Torre,Hayward,Yoon,d'Inverno2,Brady}.
The main
contribution of this paper is two formulas for the symplectic
structure of
general relativity in the double-null (2+2) formalism.
These
formulas immediately suggest two results: (i) an alternative, and
in some respects more natural
set of physical degrees of freedom based on
the twist, rather than the shear or conformal 2-metric used in most
of the previous (2+2) work, and
(ii) an alternative non-affine parametrization of
the null geodesics generating the characteristic initial-value
3-surface---in particular, ``opposite" to Hayward's suggestion
\cite{Hayward}.
By looking at the characterisitic initial-value problem we also
establish a (nonlocal) canonical transformation between the twist
and
shear type degrees of freedom.
Finally,
we also learn that
in the action principle (based on the Einstein-Hilbert action)
what is held fixed on a null spacetime boundary are the
generators of the null geodesics, up to diffeomorphisms of the
spatial 2-surfaces (and certain other subtleties which will be
addressed elsewhere \cite{Epp1}).

It should be emphasized that the symplectic structure is a very
powerful tool: classical mechanics is contained in
the statement \cite{Abraham}
\beq
i_{X_{f}}\Omega=-df,
\eeq
where $\Omega$ is the symplectic structure and $X_{f}$ is the
Hamiltonian vector field canonically generated by the observable
$f$.  In general relativity, for example, if $X_{f}$ is associated
with an asymptotic time translation of an
asymptotically flat spacetime
then $f$ is the ADM mass \cite{Ashtekar}.
In a
related, but more general context, certain diffeomorphisms are
quite nontrivial in a spacetime with boundary (such as a black
hole), and result in so-called edge degrees of freedom.  When
quantized, the latter give rise to quantum gravitational boundary
states, which may be important to providing a microscopic
understanding of black hole entropy \cite{Carlip}.  I hope to pursue
these and other related issues elsewhere \cite{Epp2}.

\vspace{1.5ex}
\begin{flushleft}
\large\bf Acknowledgements
\end{flushleft}

I am pleased to thank A. Steif, C. Torre, and especially S. Carlip
for discussions
and helpful remarks.
This work was supported by the Natural Sciences and Engineering
Research
Council of Canada, with some additional support from National Science
Foundation grant PHY-93-57203 and Department of Energy grant
DE-FG03-91ER40674.


\begin{thebibliography}{99}
\bibitem{Epp1} These and other details will be discussed more fully
in forthcoming papers by the author.
\end{thebibliography}
\end{document}